\documentclass[letterpaper,usenatbib]{mn2e}
\usepackage{epsfig}
\usepackage{natbib}

\def\apj{ApJ}

\def\apjl{ApJL}
\def\apjs{ApJS}
\def\mnras{MNRAS}
\def\aj{AJ}

\def\prd{Phys.Rev D}

\begin{document}

\title[Central and Satellite Galaxies in Cosmological SPH Simulations]{The Growth of Central and Satellite Galaxies in Cosmological Smoothed
Particle Hydrodynamics Simulations}
\author[V. Simha et al]
{Vimal Simha$^{1}$, David H. Weinberg$^{1}$, Romeel Dav\'{e}$^{2}$,
\newauthor
Oleg Y. Gnedin$^{3}$, Neal Katz$^{4}$, Du\v{s}an Kere\v{s}$^{5}$\\
$^1$ Astronomy Department and Center for Cosmology and AstroParticle Physics, Ohio State University, Columbus, OH 43210,\\
vsimha,dhw@astronomy.ohio-state.edu\\
$^2$University of Arizona, Steward Observatory, Tuscon, AZ 85721,
rad@as.arizona.edu\\
$^3$Department of Astronomy, University of Michigan, MI 48109,
ognedin@umich.edu\\
$^4$Astronomy Department, University of Massachusetts at Amherst, MA 01003, nsk@kaka.phast.umass.edu\\
$^5$Institute for Theory and Computation, Harvard-Smithsonian Center for
Astrophysics, Cambridge, MA 02138, dkeres@cfa.harvard.edu\\
}

\maketitle
\begin{abstract}

We examine the accretion and merger histories of central and satellite 
galaxies in a smoothed particle hydrodynamics (SPH) cosmological simulation
that resolves galaxies down to $7\times 10^9 M_\odot$.  Most friends-of-friends
haloes in the simulation have a distinct central galaxy, typically two to five
times more massive than the most massive satellite.  As expected, satellites
have systematically higher assembly redshifts than central galaxies of the
same baryonic mass, and satellites in more massive haloes form earlier.  
However, contrary to the simplest expectations, satellite galaxies continue to
accrete gas and convert it to stars; the gas accretion declines steadily over
a period of $0.5-1$ Gyr after the satellite halo merges with a larger parent
halo.  Satellites in a cluster mass halo eventually begin to lose baryonic
mass.  Since $z=1$, 27\% of central galaxies (above $3\times 10^{10} M_\odot$)
and 22\% of present-day satellite galaxies have merged with a smaller system
above a 1:4 mass ratio; about half of the satellite mergers occurred after
the galaxy became a satellite and half before.  In effect, satellite galaxies
can remain ``central'' objects of halo substructures, with continuing
accretion and mergers, making the transition in assembly histories and
physical properties a gradual one.  Implementing such a gradual transformation
in semi-analytic models would improve their agreement with observed colour
distributions of satellite galaxies in groups and with the observed colour dependence
of galaxy clustering.

\end{abstract}


%\newpage

\section{Introduction}
%\label{sec:intro}

In the standard theoretical description of galaxy formation, galaxies
grow initially by the condensation of gas at the centres of dark
matter potential wells \citep[e.g.][]{white78,fall80}. When a large
dark matter halo accretes a smaller halo, the galaxy hosted by the
smaller halo becomes a satellite system, which may eventually merge
with the central object after sinking by dynamical friction. Even if
the satellite retains some of its original dark matter, it is likely
to move through the large parent halo with a random velocity that
exceeds its own escape speed, making it difficult for the satellite to
accrete gas from the halo or to merge with other satellites. In
semi-analytic models of galaxy formation \citep[e.g.][]{white91,kauffmann93,cole94,somerville98,croton06,bower06},
these expected differences between central and satellite galaxies are
usually encoded by simple rules, with some variations from one model
to another. When two haloes merge, the central galaxy of the more
massive progenitor becomes the central galaxy of the new, larger
halo. Cooling gas from the halo is assigned mostly or entirely to the
central galaxy. Satellites accrete little or no fresh gas, and their
pre-existing gas may be removed by ram pressure stripping when they
enter the larger halo. Most or all mergers are assumed to involve
satellites joining the central galaxy rather than merging with each
other.

In this paper, we examine the growth of central and satellite galaxies
in smoothed particle hydrodynamics (SPH) cosmological simulations
\citep{katz92b,evrard94,katz96}. In such simulations, differences
between central and satellite galaxies are not imposed a priori, but
central galaxies nevertheless emerge as a distinct class for the
physical reasons discussed above. The galaxy closest to the halo
centre of mass is usually older and more massive than satellites in
the same halo \citep{berlind03,zheng05}. Here we investigate the
assembly histories of present day centrals and satellites as a
function of galaxy and halo mass, and we track the growth of galaxies
to see the extent to which satellite systems experience continuing gas
accretion or mergers with other satellites.

Our goals are to understand the physics of galaxy assembly in the
simulation and to inform semi-analytic models of galaxy formation and
other efforts to interpret observed galaxy clustering and the
environmental dependence of galaxy properties. The central-satellite
distinction is a key element in modeling the origin of bimodality in
the galaxy population, in which galaxies with old stellar populations
and little ongoing star formation form a distinct, passive, ``red
sequence'' \citep{strateva01,blanton01,bell03,faber07}. Strangulation
of gas supplies in satellites is the likely route by which many
galaxies join the passive sequence, though additional mechanisms are
required to explain the red colours and low star formation rates of
the most massive galaxies \citep[see e.g.][]{croton06,dekel06,cat07}. Central-satellite separation is also a
key ingredient in ``halo occupation distribution'' (HOD) models of
galaxy clustering \citep{kravtsov04,zheng05} and in recent
efforts to interpret properties of galaxy groups \citep[e.g.][]{weinemann06,gilbank07,hansen07,cli07,coil08}.

The simulation that we analyse here is also the main simulation
analysed by \cite{keres05} and \cite{maller06}, who examine gas
accretion and merger rates but do not distinguish between central and
satellite galaxies. \cite{keres05} emphasise the distinction between
between ``cold'' and ``hot'' modes of gas accretion. In haloes of total
mass $M\le3 \times 10^{11} M_{\odot}$, most gas that accretes onto
galaxies does so without ever heating close to the halo virial
temperature. In higher mass haloes, most gas heats to the virial
temperature before settling into galaxies. Overall, about half of the
baryonic mass in the simulation's $z=0$ galaxy population was
originally accreted in ``cold mode'' and half in ``hot mode'', with
the former dominating at high redshift and in low mass galaxies
today. Here we will investigate both total and cold-only accretion
rates of central and satellite systems. Further discussions of the
cold/hot dichotomy appear in
\cite{binney77}, \cite{binney03}, \cite{katz92a}, \cite{kay00}, \cite{fardal01}, \cite{katz03}, \cite{birnboim03}, \cite{croton06}, 
\cite{dekel06} and \cite{keresp}.

We describe our simulation and methods for identifying haloes and
galaxies in \S2, before proceeding to our main results in \S3. There
we start with a qualitative overview of accretion and merger
properties of central and satellite galaxies, then quantify their
relative accretion rates and merger rates and their dependences on
halo mass. In \S4 we summarise our results and discuss their
implications, with particular attention to recent studies of galaxy
bimodality and the red population in groups.

\section{Simulation and Numerical Methods}
\label{sec:simualtion}

\subsection{Simulation}

Our simulation is performed using a parallel implementation of TreeSPH
\citep{dave97,hernquist89,katz96}, which combines a tree algorithm for
gravitational calculations with smoothed particle hydrodynamics
(SPH). There are three kinds of particles in our simulation: dark
matter, stars and gas. The collisionless particles (dark matter and
stars) are influenced only by gravity, whereas the gas particles are
influenced by pressure gradients and shocks in addition to
gravity. Gas particles experience adiabatic heating and cooling, shock
heating, inverse Compton cooling off the CMB and radiative cooling via
free-free emission, collisional ionisation and recombination and
collisionally excited line cooling. Further details of the code can be
found in \cite{katz96}.

We model a comoving $22.222$h$^{-1}$ Mpc periodic box with $128^3$ gas
particles and $128^3$ dark matter particles. The dark matter particle
mass is 7.9 $\times$ $10^8$ M$_{\odot}$, and the SPH particle mass is
1.05 $\times$ $10^8$ M$_{\odot}$. The gravitational force softening is
a comoving 5$h^{-1}$ kpc cubic spline, which is roughly equivalent to a
Plummer force softening of 3.5$h^{-1}$ kpc.

We include star formation as described by \citet{katz96}. Overdense,
Jeans unstable gas that is part of a converging flow with density
$\rho_{\rm gas} > 0.1 m_{\rm H} {\rm cm}^{-3}$ and temperature $T \leq
30,000$K is converted to stars on a timescale set by the dynamical
timescale or the cooling time, whichever is longer. Our prescription
for star formation leads to a relation with gas surface density
similar to a Schmidt law. We adopt a \cite{miller79} initial mass
function. We include supernova feedback with $7.35 \times 10^{-3}$
supernovae per solar mass, and each supernova deposits $10^{51}$ ergs
of energy into the surrounding medium. Since the surrounding gas is
dense, this energy is usually radiated away before it can drive
outflows or suppress subsequent star formation.

We adopt a $\Lambda$CDM cosmology (inflationary cold dark matter with
a cosmological constant) with $\Omega_m$=0.4, $\Omega_{\Lambda}$=0.6,
$h$=0.65, $\Omega_b=0.02h^{-2}$, spectral index $n$=0.93, and
$\sigma_8$=0.8. These values are reasonably close to the estimates
from the cosmic microwave background \citep{dunkley08} and large scale
structure \citep{tegmark06}, though these favour somewhat lower
$\Omega_m$ ($\approx 0.25$) and higher $h$ ($\approx 0.7$). We do not
expect modest differences in cosmological parameters to influence our
conclusions.

Hydrodynamic cosmological simulations produce dense groups of baryons
with sizes and masses comparable to the luminous regions of observed
galaxies \citep{katz92a}. We identify galaxies using the Spline Kernel
Interpolative DENMAX (SKID) algorithm \citep{gelb94,katz96}, which
identifies gravitationally bound particles associated with a common
density maximum. We apply this algorithm to all baryonic particles
with $T < 3 \times 10^4K$ and $\rho_{\rm gas}/\overline{\rho}_{\rm
gas} > 10^3$, and we refer to the groups of stars and cold gas thus
identified as galaxies. The simulated galaxy population becomes
substantially incomplete below our resolution threshold of 64 SPH
particles \citep{murali02}, which corresponds to a baryonic mass of
$6.8 \times 10^9$ M$_{\odot}$. We ignore lower mass galaxies in our
analysis. There are 1120 galaxies above the resolution threshold in
our simulation.

We identify dark matter haloes using a FOF (friends-of-friends) algorithm
\citep{davis85}. The algorithm selects groups of particles in which
each particle has at least one neighbour within a linking length. We
adopt a linking length $l=0.2 \overline {n}^{-1/3}$, where
$\overline{n}$ is the mean dark matter particle density, which
leads to haloes with typical mean overdensity $\rho / \overline{\rho}
\approx 200$. By chance, the simulation contains one halo that is
anomalously massive ($M_h = 3.4 \times 10^{14} M_{\odot}$) for the 
simulation volume, which has some impact on our statistics for 
satellite galaxies when these are not broken down into halo mass
bins.

We use $227$ output epochs between $z=9$ and $z=0$, with a typical
spacing between output epochs of less than 150 Myr. At each epoch, we
identify haloes and galaxies above the resolution threshold.

\subsection{Central and satellite galaxies}

\begin{figure}
\centerline{
\epsfxsize=3.7truein
%\epsfbox[41 17 570 733]{figure1.eps}
\epsfbox{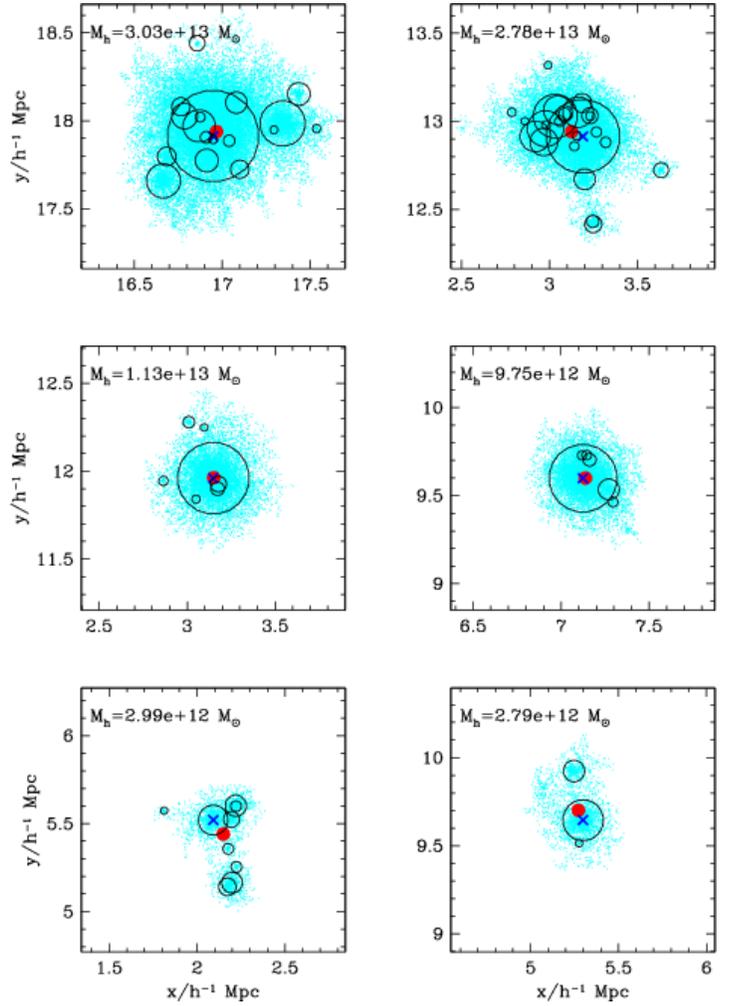}
}
\caption{
A selection of friends-of-friends haloes and their galaxy populations. The small
blue dots are dark matter particles, the galaxies are shown as open black
circles with their areas proportional to their masses. Each panel is centred on
the most bound dark matter particle shown as a blue cross, and the centre of mass of the dark matter
particles is shown as a red filled circle.  
}
\label{fig:haloes}
\end{figure}

Figure \ref{fig:haloes} shows a selection of six FOF haloes and their
galaxy populations at $z=0$. Each panel is centred on the most bound
dark matter particle in the halo. The centre of mass is shown as a red
filled circle in the figure. The galaxies are shown as open circles
with areas proportional to their (baryonic) mass. A large majority of
haloes have a clear central galaxy that is more massive than other
galaxies in the halo and located at the bottom of the dark matter
potential well. However, in a small number of haloes there are two
galaxies of comparable mass, each with its own satellites. Some of the
systems without a clear central galaxy consist of distinct
sub-groups that are in the process of undergoing a merger, as in the
lower left panel of Figure \ref{fig:haloes}.

We hereafter define the central galaxy of each halo to be the galaxy
with the largest baryonic mass. Inspection of many haloes shows that
this identification is always reasonable when the halo is regular, and
as reasonable as any other choice for the small fraction of irregular
haloes.

\begin{figure}
\centerline{
\epsfxsize=3.7truein
%\epsfbox[49 484 562 680]{figure2.eps}
\epsfbox{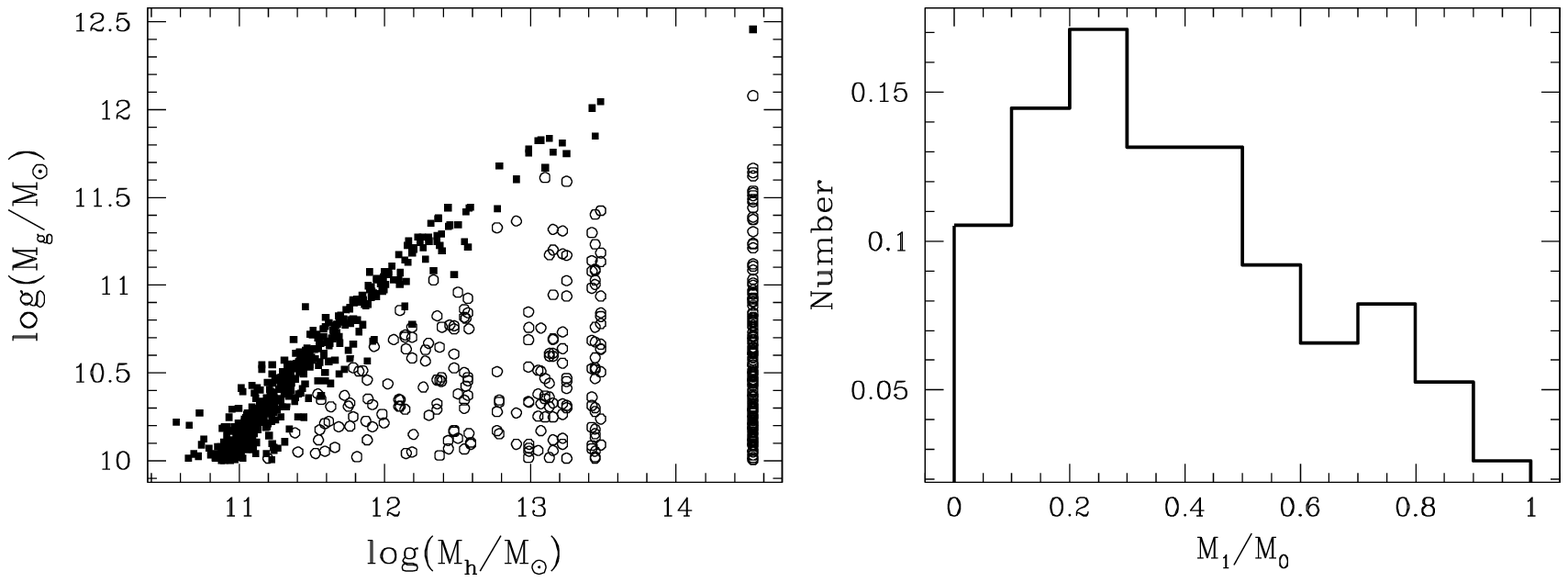}
}
\caption{
}\textit{(Left)} Masses of central galaxies (squares) and satellite galaxies
(circles) plotted against halo mass.
\textit{(Right)} A histogram of the ratio of the mass M$_1$ of the most massive
satellite galaxy to the mass M$_0$ of the central galaxy. 
\label{fig:sub0}
\end{figure}

Figure \ref{fig:sub0} shows the masses of galaxies and haloes in the
simulation.  The left hand panel plots galaxy mass versus halo mass,
with central galaxies shown as filled squares and satellite galaxies
shown as open circles.  The right hand panel plots the distribution of $M_1
/ M_0$, where $M_0$ is the mass of the central galaxy and $M_1$ is the
mass of the most massive satellite. We omit haloes that have no
satellites above the resolution threshold. The distribution of
satellite mass ratios is broad, and more than two-thirds of haloes have
$0.2 \le M_1/M_0 \le 0.5$. Only a small fraction of haloes have two
galaxies of comparable mass ($M_1/M_0 \ge 0.8$).

\begin{onecolumn}
\begin{figure}
\centerline{
\epsfxsize=7.5truein
\epsfbox{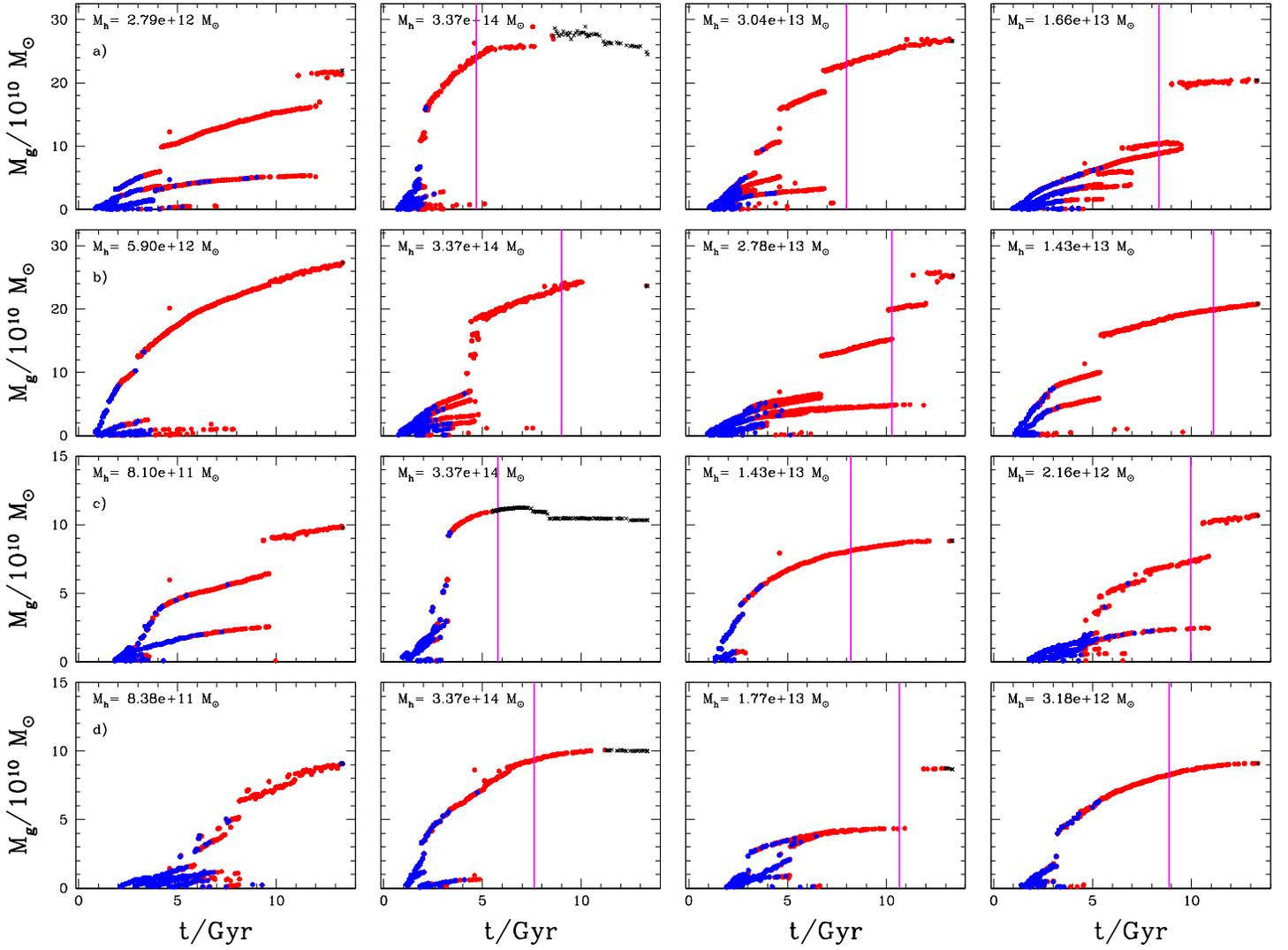}
}
%\label{fig:sub1}
\caption{
Evolution of galaxies with time. Cold accretion is shown in blue and hot
accretion in red. $M_h$ (top left corner) is the halo mass. The black crosses
show the subsequent evolution of the mass of the galaxy after its last accretion
event. For the satellite galaxies, the magenta coloured vertical line marks the
epoch at which it became a satellite.
The first column shows central galaxies, while columns $2$, $3$ and $4$ show
satellite galaxies in decreasing order of halo mass. 
Panels $(a)$ and $(b)$ show galaxies with masses of $\sim 3 \times 10^{11}
M_{\odot}$ while panels $(c)$ and $(d)$ show galaxies with masses of $\sim
10^{11} M_{\odot}$. 
}
\label{fig:sub1}
\end{figure}

\begin{figure}
\centerline{
\epsfxsize=7.5truein
\epsfbox{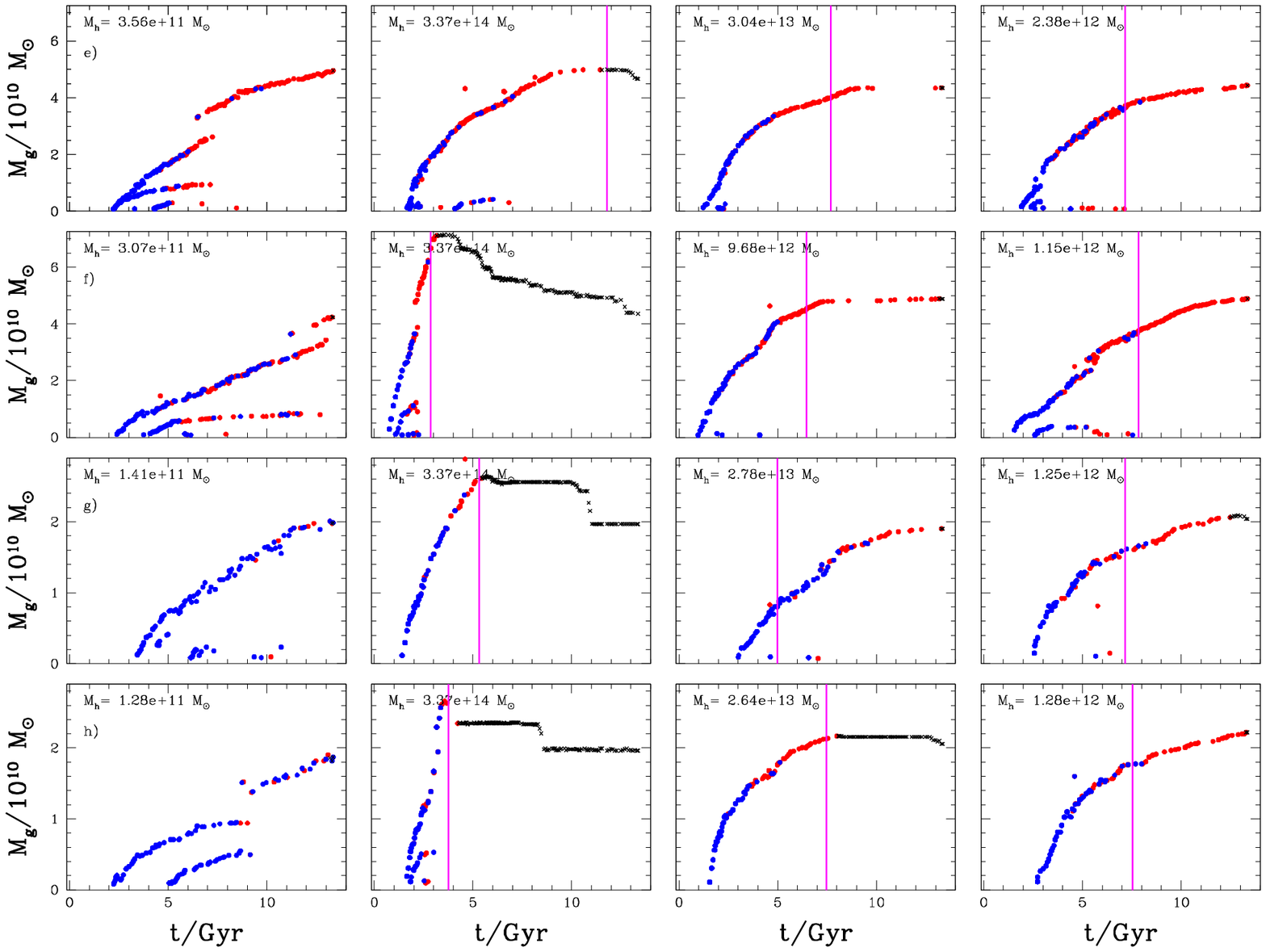}
}
Figure~\ref{fig:sub1}(continued)
Panels $(e)$ and $(f)$ show galaxies with masses of $\sim 5 \times 10^{10}
M_{\odot}$ while panels $(g)$ and $(h)$ show galaxies with masses of $\sim 2
\times 10^{10} M_{\odot}$. 
\end{figure}

\begin{figure}
\centerline{
\epsfxsize=5.0truein
\epsfbox{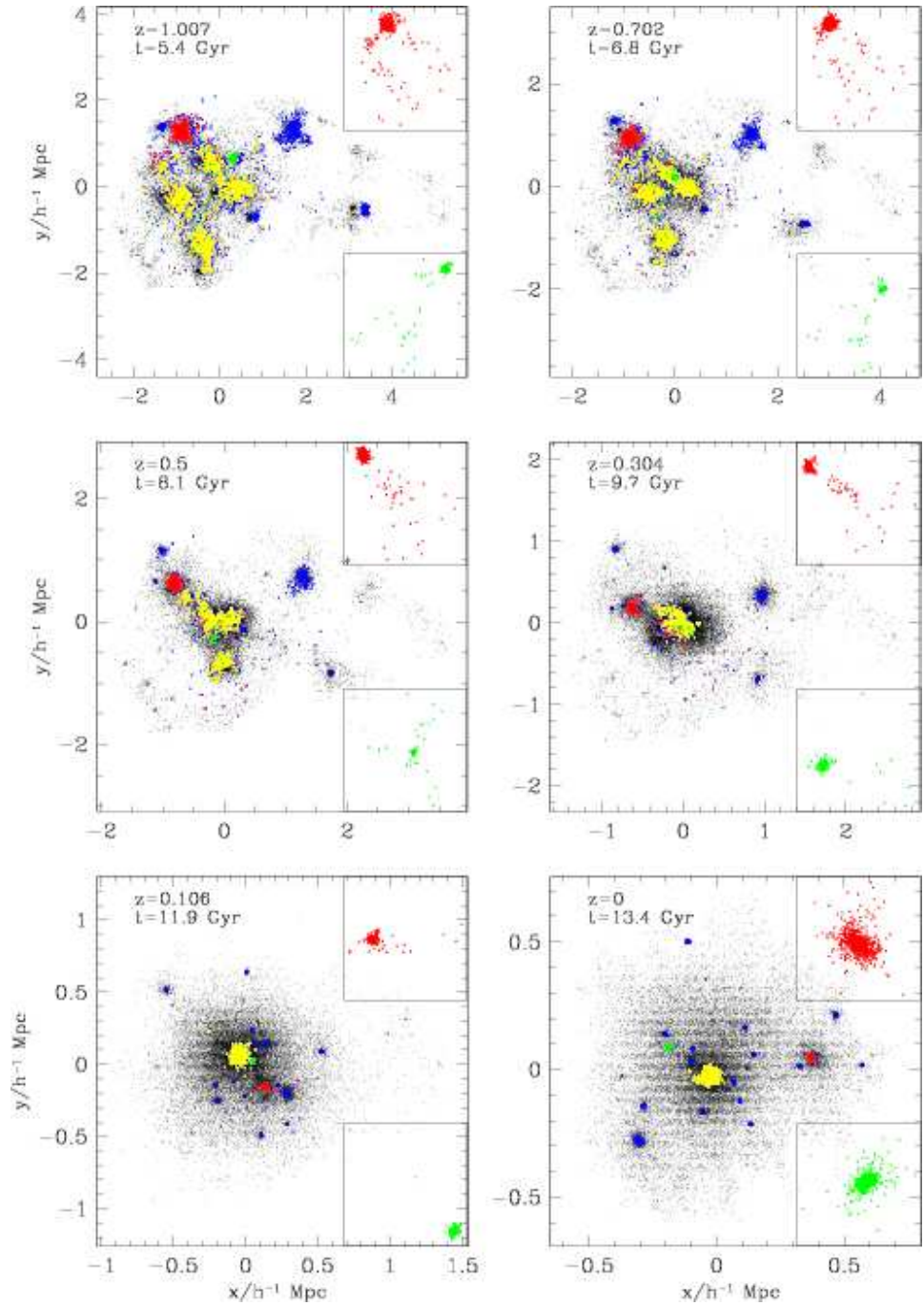}
}
\caption{
The evolution of a $~3\times 10^{13} M_{\odot}$ halo with redshift, with
positions marked in comoving $h^{-1}$ Mpc. Dark matter particles are shown as
black dots. The particles in the central galaxy at $z=0$ are shown in yellow.
The red galaxy is the most massive satellite galaxy, which becomes a satellite
at $z=0.514$, while the green galaxy becomes a satellite at $z=0.736$. The
baryonic particles of other $z=0$ satellites are shown in blue. The inset panels in
the right hand corners show a magnification of the red and green satellites,
large enough to encompass all of their member particles. While the central galaxy
has a large total extent at $z=0$, its stellar half-mass radius is only 3.9 $h^{-1}$ kpc.
}
\label{hhisplot}
\end{figure}

\end{onecolumn}

\twocolumn

\section{ACCRETION AND MERGER PROPERTIES: CENTRAL AND SATELLITE GALAXIES}

Figure \ref{fig:sub1} shows the assembly history of a representative
sample of central and satellite galaxies. At each of our 227 output
epochs from the simulation, we use SKID to identify galaxies with at
least 8 SPH particles. For galaxies resolved in two consecutive output
epochs, we identify gas particles that are in the galaxy at the later
epoch but are not in any resolved galaxy at the earlier epoch as
smoothly accreted gas. In Figure \ref{fig:sub1}, galaxy mass is shown
on the vertical axis, and time is shown on the horizontal axis.  Each
point represents an accretion event, showing the baryonic mass of the
SKID identified galaxy when the particle was accreted onto it. Every
particle that is in the $z=0$ group is shown. The locus traced by the
topmost points shows the growth of the most massive progenitor of the
$z=0$ galaxy, while lower ridges trace the growth of secondary
progenitors which merge onto the main progenitor (or with each other)
producing jumps in the mass track. There are occasional outlier
points, cases in which an accreting particle is incorrectly assigned
to a more massive neighboring system instead of the galaxy in which it
is actually located.

Following \cite{keres05}, we distinguish between cold and hot
accretion. We trace the temperature history of each particle and
identify the maximum temperature $T_{\rm{max}}$ that the particle had
prior to accreting onto a galaxy since the beginning of the
simulation. Particles with $T_{\rm{max}} \leq 2.5\times10^5$K are
classified as cold and shown in blue colour, and particles with
$T_{\rm{max}} \geq 2.5\times10^5$K are classified as hot and shown in
red. Since each point corresponds to accretion of one particle, intermediate
mass galaxies sometimes show interleaved hot and cold accretion events.

For satellite galaxies, we trace the history of its most massive
progenitor galaxy and identify the epoch when it became a satellite,
which we define as the epoch when the parent halo of the satellite
merges with the parent halo of a more massive galaxy. In Figure
\ref{fig:sub1}, the magenta coloured vertical line marks the epoch when
the galaxy became a satellite.

Rows (a) and (b) of Figure \ref{fig:sub1} show high mass galaxies with
final baryonic masses (consisting of stars and cold, dense gas) of
$M_g \sim 2-3 \times 10^{11} M_\odot$. Subsequent rows show galaxies
with final baryonic masses $M_g \sim 10^{11} M_\odot$ (c and d),
$M_g \sim 4-5 \times 10^{10} M_\odot$ (e and f) and $M_g \sim 1-2
\times 10^{10} M_\odot$ (g and h). In each row, the first column
shows a central galaxy in a halo of mass $M_h \sim 5-10$ $M_g$.
Subsequent columns show satellite galaxies in decreasing order of halo
mass. The second column shows satellite galaxies selected from the
most massive halo in the simulation ($M_h =
3.4\times10^{14}M_{\odot}$). The third and fourth columns show
satellite galaxies from intermediate mass haloes and low mass haloes
respectively. While these thirty-two examples cannot capture the full
range of assembly histories that we see in the simulation, they
suffice to illustrate the main qualitative trends.

An examination of Figure \ref{fig:sub1} reveals a number of
interesting trends. The assembly of high mass galaxies is complex,
with active merger histories and continuing accretion at late
times. Among high mass galaxies, there is not much difference between
the assembly histories of central galaxies and satellite galaxies,
except for the satellites in the most massive halo. Satellites in
intermediate mass haloes, as shown in columns 3 and 4, continue to
accrete gas and form stars after $z_{\rm{sat}}$, the epoch when they
become satellites. Some systems, such as the galaxy in the third
column of row (b) and the fourth column of rows (a) and (c),
``receive'' mergers after becoming satellites. However, satellites in
the highest mass halo are an exception. For these, assembly occurs
early and there is little accretion after $z_{\rm{sat}}$, and there is
mass loss in some cases such as row (c) column 2.

The assembly of lower mass galaxies appears to be smoother with fewer
merger events.  However, because our galaxy population is incomplete
below our resolution threshold of $6.8\times10^9 M_{\odot}$, it is
likely that we miss many lower mass merger events. Lower mass galaxies
in low and intermediate mass haloes often have continuing accretion and
star formation after becoming a satellite, with row (g), column 3
being a striking example. Some lower mass galaxies in intermediate
mass haloes, such as column 3 of rows (e) and (f), have low but
non-zero accretion rates at late times. The assembly histories of low
mass galaxies in the most massive halo are systematically different
from their central counterparts. They experience very little accretion
after becoming satellites, and in many cases they experience
significant mass loss (column 2 of rows f-h).

Overall, Figure \ref{fig:sub1} shows surprisingly little distinction
between central and satellite galaxies. Galaxy mass is generally a
stronger predictor of growth history than central/satellite status. We
do see some trends of the expected sign, but $z_{\rm{sat}}$, when the
galaxy becomes a satellite, is not a sharp boundary in the accretion
or merger history.  Satellites in the most massive halo are an
exception, in that accretion is slowed or shut-off at $z_{\rm sat}$,
and some satellites experience mass loss.  We have examined assembly
histories of many more systems in addition to the 32 shown in Figure
\ref{fig:sub1}, and they confirm the qualitative points made here.

\begin{figure}
\centerline{
\epsfxsize=3.5truein
\epsfbox{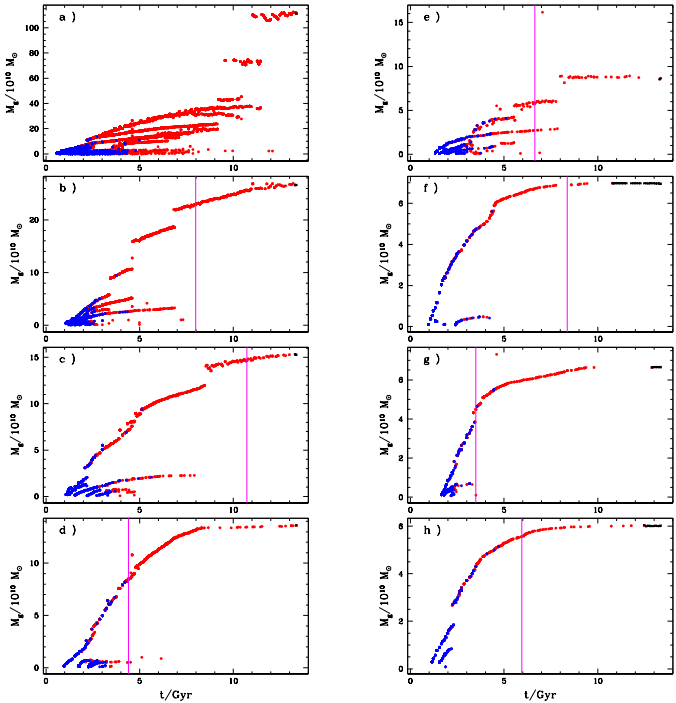}
}
\caption{
Galaxy mass vs time for galaxies in the halo of Figure~\ref{hhisplot}. Panel $(a)$ is
the central galaxy, $(b)$ is the most massive satellite, shown in red in
Figure~\ref{hhisplot}  and $(e)$ is the satellite shown in green in
Figure~\ref{hhisplot}. The format is the same as figure~\ref{fig:sub1}
}
\label{haloaccplot}
\end{figure}

It is remarkable that satellites in low and intermediate mass haloes,
which are moving at high velocities in the potential of the halo,
continue to accrete gas and form stars. Our conjecture is that
continuing accretion by satellites owes to substructure in the
haloes. In effect, a satellite galaxy in a large halo may still be the
central galaxy of its own coherent sub-group. Figure \ref{hhisplot}
investigates one such system, a $3 \times 10^{13} M_{\odot}$ halo as
it evolves from $z=1.07$ to $z=0$. We plot the dark matter particles
in black and the baryonic particles in different colours, which depend
on their galaxy membership at $z=0$. Figure \ref{haloaccplot} shows
the corresponding assembly histories of the eight most massive
galaxies in this halo. Consider the most massive satellite galaxy,
shown in red in Figure \ref{hhisplot} (and in panel b of Figure
\ref{haloaccplot}). It starts off as a central galaxy in its own
halo. Following the merger of its parent halo with the parent halo of
the central galaxy at $z=0.514$, it continues to accrete gas from the
region around it and has other galaxies gravitationally bound to it,
in effect its own satellites. Even at $z=0$, it is the central object
of a distinctly defined dark matter substructure. Hence, accretion is
not shut off at $z_{\rm{sat}}$, and there is no sharp transition in
its behaviour at this redshift. Another satellite galaxy, shown in
green in Figure \ref{hhisplot}, becomes a satellite at $z=0.736$, but
it receives a merger after that time.

\begin{figure}
\centerline{
\epsfxsize=3.5truein
\epsfbox{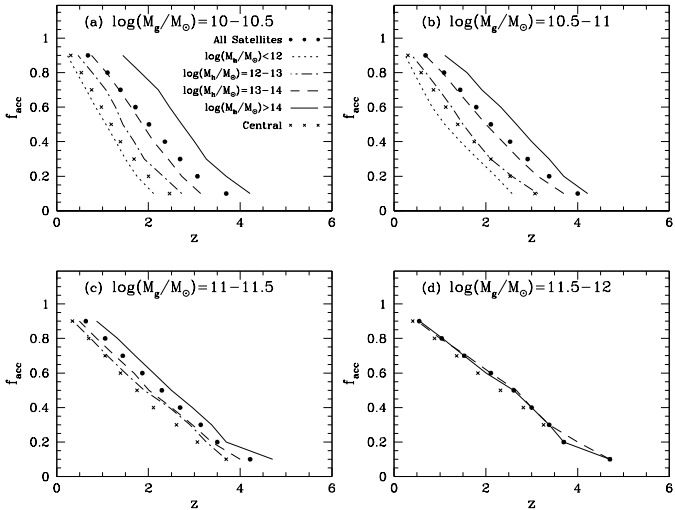}
}
\caption{
Fraction of final mass accreted by redshift $z$ is plotted against redshift. We
include mass in all progenitors of the $z=0$ system. In each panel,
crosses show accretion histories of central galaxies, circles show all
satellites, and lines show satellites in bins of halo mass as indicated in the
legend of panel (a).
}
\label{fig:sub3}
\end{figure}

We now turn to statistical measures to quantify the anecdotal results
seen in Figures 3-5. Figure \ref{fig:sub3} compares the assembly
histories of central and satellite galaxies as a function of galaxy
mass and parent halo mass. For four galaxy mass bins, we plot the mass accreted by
redshift $z$ onto any of a galaxy's SKID-identified progenitors divided by the total mass accreted up to $z=0$. 
In each panel, the crosses show the
results for all the central galaxies, the circles show all the
satellite galaxies, and the lines show the satellites in different
bins of halo mass.

Looking first at central galaxies, there is a clear trend for higher
mass galaxies to assemble at higher redshifts. For example, the
redshift at which 50\% of the final galaxy mass has been accreted by
the central galaxy population ($f_{\rm acc}$=0.5 in Figure
\ref{fig:sub3}) falls from $z=2.2$ for $M_{\rm g} =
10^{11.5}$M$_{\odot}$-10$^{12}$M$_{\odot}$ to $z=1.2$ for $M_{\rm g} =
10^{10.5}$M$_{\odot}$-10$^{11}$M$_{\odot}$, with a steady trend in
between these masses. This downward shift of assembly redshifts is
consistent with the well known ``downsizing'' trend of observed galaxy
evolution. It arises naturally: massive galaxies reside in dense
environments with accelerated early formation \citep{neistein06}. However, even the most
massive galaxies have a significant fraction of mass accreted at low
redshift, which is inconsistent with the observed red colours of the
most massive galaxies.

Low mass satellite galaxies accrete most of their mass early on and
experience little accretion at late times, with $f_{\rm acc}$=0.5 at
$z=2$ compared to $z=1.3$ for low mass centrals. There is a strong
trend with halo mass: low mass satellites in the most massive halo
assemble earlier and have accreted 90\% of their final mass by
$z=1.5$, while the satellites in $M_{\rm h}\le10^{12}$M$_{\odot}$
haloes have slightly lower assembly redshifts than comparable central
galaxies. The same trends are seen for higher galaxy masses, but the
differences between central and satellite galaxies are smaller. For
high galaxy masses (panels c and d), there are no satellites in low
mass haloes.

With our star formation prescription, which results in an increasing
star formation rate with increasing gas density, star formation
closely tracks the accretion of gas with a time lag
\citep[see][]{keresp}. We find similar trends for star formation as we did
for gas accretion: satellite galaxies typically form their stars
earlier than central galaxies, but the difference is smaller at higher
galaxy masses. Among satellites, those in high mass haloes typically
form stars early on and have little star formation at late times,
while satellites in low mass haloes continue to form stars at low
redshifts.

To further investigate the physical processes involved in gas
accretion by satellites, we differentiate between hot and cold
accretion \citep{keres05}. There are two reasons that it is
interesting to isolate the contribution of cold accretion to galaxy
growth. First, tests by \cite{keres07} and \cite{keresp} show that the
rate of hot gas accretion in hydrodynamic cosmological simulations is
sensitive to differences among simulation codes but the cold accretion
rates are robust. In particular, hot accretion rates are lower in
simulations with GADGET \citep{springel01} than in simulations with
the PTreeSPH code used here, probably because the former employs an
entropy conserving formulation of SPH. Second, AGN feedback, which is
not incorporated in our simulation, could be more effective at
suppressing hot accretion flows, since they have low-densities and are
quasi-spherical compared to cold accretion flows, which channel dense
gas along filaments \citep{keres05}. Thus, for either numerical or
physical reasons, the cold accretion rate in our simulation might be a
better approximation to a galaxy's true total accretion rate.

\begin{figure}
\centerline{
\epsfxsize=3.5truein
\epsfbox{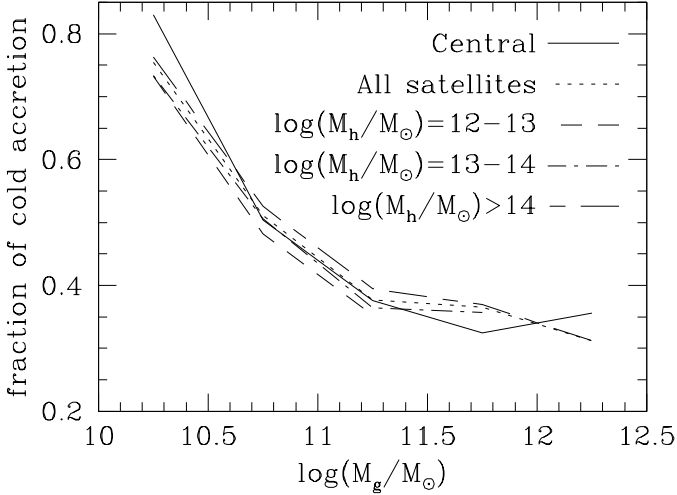}
}
\caption{
The fraction of $z=0$ galaxy mass that was gained through cold accretion, as a
function of galaxy mass. The solid line shows central galaxies, the dotted line
shows all satellite galaxies and other lines show satellite galaxies in bins of
halo mass as indicated in the legend.
}
\label{fig:sub5}
\end{figure}

Figure \ref{fig:sub5} shows the fraction of $z=0$ galaxy mass that was
originally acquired (by the galaxy or its progenitors) via cold
accretion. In agreement with \cite{keres05}, we find that low mass
galaxies gain their mass almost entirely from cold accretion, while
the highest mass galaxies gain up to ~2/3 of their final mass by hot
accretion. (This latter behaviour is different in the Gadget-2
simulations of \citeauthor{keresp}[\citeyear{keresp}], but the former remains true; see their Figure 10.)  Figure
\ref{fig:sub5} shows that the trend with galaxy mass is nearly
identical for central and satellite galaxies, and that among satellite
galaxies there is no dependence of the cold accretion fraction on halo
mass. If hot accretion were completely suppressed in the real universe
(or in numerical simulations with a different treatment of gas physics
or feedback), then to a first approximation the $z=0$ galaxy masses would be reduced, on
average, by the factor shown in Figure \ref{fig:sub5}.

\begin{figure}
\centerline{
\epsfxsize=3.5truein
\epsfbox{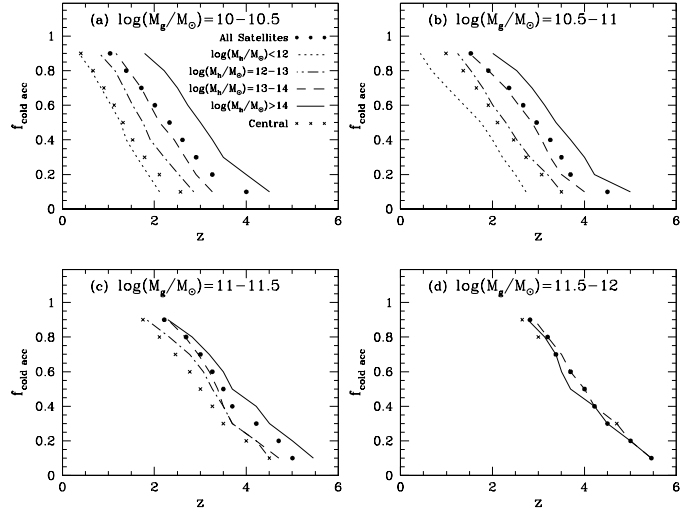}
}
\caption{
Fraction of mass accreted in the cold mode vs redshift. This figure is analogous
to Figure \ref{fig:sub3}, but we eliminate all material accumulated through hot
accretion.
}
\label{fig:sub3a}
\end{figure}

Figure \ref{fig:sub3a} repeats the assembly history analysis of Figure
\ref{fig:sub3}, but we now eliminate all material gained originally
through hot accretion (onto the galaxy or any of its progenitors)
before computing galaxy masses at any redshift. All of the trends seen
in Figure \ref{fig:sub3} are still apparent: assembly shifts towards
higher redshifts for higher galaxy masses, satellite galaxies assemble
earlier than central galaxies of the same mass, and satellites in
massive haloes assemble earlier than satellites in low mass
haloes. However, the trend of assembly history with galaxy mass is much
stronger (for centrals and satellites) because galaxies experience
little cold accretion once their mass exceeds $M_{g}\sim 3 \times
10^{10}$M$_{\odot}$ \citep{keres05}. Galaxies above $M_{
g}=10^{11}$M$_{\odot}$ have typically accreted 90\% of their final
mass by $z=2$. Suppressing hot accretion is thus a way of turning the
most massive galaxies ``red and dead''
\citep{binney03,croton05,keres05,dekel06,cat07}.

So far, we have compared the properties of central galaxies to the
properties of satellite galaxies over the course of their lifetime,
from the time of their formation until $z=0$. However, these
satellites were once central galaxies, and they only become satellites
at the redshift $z_{\rm sat}$, when their parent halo merged with the
parent halo of a more massive galaxy. To fairly compare the
post-$z_{\rm sat}$ growth of satellites to central galaxies, we create
a ``control'' sample. We associate each satellite galaxy above the
resolution threshold with a randomly chosen central galaxy of similar
baryonic mass. We follow the evolution of this matched central galaxy
from $z_{\rm sat}$ of the corresponding satellite galaxy to
$z=0$. Thus, the control sample has the same mass distribution as the
satellites and is traced over the same range of redshifts.

\begin{figure}
\centerline{
\epsfxsize=3.5truein
\epsfbox{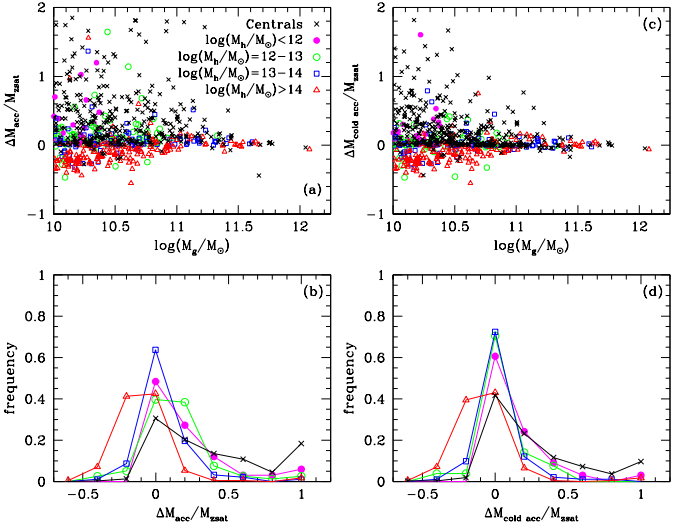}
}
\caption{
\textit{(Left)} Mass change from accretion after becoming a satellite as a
function of galaxy mass. Each point represents a galaxy. The red triangles, blue
squares, green open circles and magenta filled circles show satellites in haloes
of different masses. Each satellite galaxy is paired with a central galaxy of
similar mass which is shown as a cross. $M_g$ is the mass of the galaxy at
$z=0$. $M_{\rm zsat}$ is the mass of all the progenitors of the galaxy at the epoch
when the most massive progenitor became a satellite. Panel (b) shows a histogram
of the fractional mass change from accretion for satellites in different halo
mass bins as well as central galaxies. (To preserve visual clarity, we plot this
histogram as connected points.)
\textit{(Right)} Mass change from cold accretion after becoming a satellite.
Panel (d) shows a histogram of the fractional mass change from cold accretion
for satellites in different halo mass bins as well as central galaxies.  
}
\label{fig:sub6}
\end{figure}

Figure \ref{fig:sub6} compares the mass accreted by satellite galaxies
after they become satellites to the mass accreted by the control
sample of central galaxies over the same time period. Panel (a) shows
the mass accreted by each satellite galaxy from $z_{sat}$ to $z=0$ as
a fraction of its mass at $z_{sat}$. Satellite galaxies in four halo
mass bins are represented by different colours and point shapes. For
comparison, we show the mass accreted by our control sample of central
galaxies over the same time interval. Central galaxies tend to gain
mass after $z_{sat}$ of their matched satellites, with some even
tripling their mass between $z_{sat}$ and $z=0$. For satellites, there
is a trend with halo mass. Satellites in low mass haloes ($M_h \leq
10^{12} M_{\odot}$) gain a small fraction of their mass after
$z_{sat}$.  Satellites in intermediate mass haloes
($M_h=10^{12}-10^{14} M_{\odot}$) tend neither to lose nor gain much
mass; their mass at $z=0$ is similar to their mass at $z_{\rm
sat}$. Satellites in the high mass, $M_h = 4 \times 10^{14}
$M$_{\odot}$ halo typically lose mass, in some cases up to half the
mass they had at $z_{\rm sat}$. The continuous nature of the trend
with halo mass is more evident in panel (b), which presents histograms
of the accreted mass fractions summed over all galaxy mass bins.

Panel (c) shows the mass accreted by galaxies in the cold mode from
$z_{\rm sat}$ to $z=0$, again scaled to the mass at $z_{\rm sat}$
(which still includes hot mode accretion up to that epoch). As in
panel (a), we show satellite galaxies differentiated by halo mass, and
we use the same control sample of central galaxies. For low mass
galaxies, most of the accretion takes place in the cold mode. Hence,
for low mass galaxies, panel (c) is identical to panel (a). However,
for higher mass galaxies most of the accretion takes place through the
hot mode, so above $M_{g} \sim 10^{10.5} $M$_{\odot}$ the
accretion rates are lower for both central and satellite galaxies. As
before, panel (d) presents histograms summed over bins of galaxy
mass. It is again clear that central galaxies tend to gain mass after
the matched $z_{\rm sat}$ and that satellites in the most massive halo
tend to lose mass. However, with the lowered satellite accretion
rates, the continuous trend in intermediate mass haloes is no longer so
clear - there is only a small amount of satellite accretion in the
cold mode, even in relatively low mass haloes.

\begin{figure}
\centerline{
\epsfxsize=3.5truein
\epsfbox{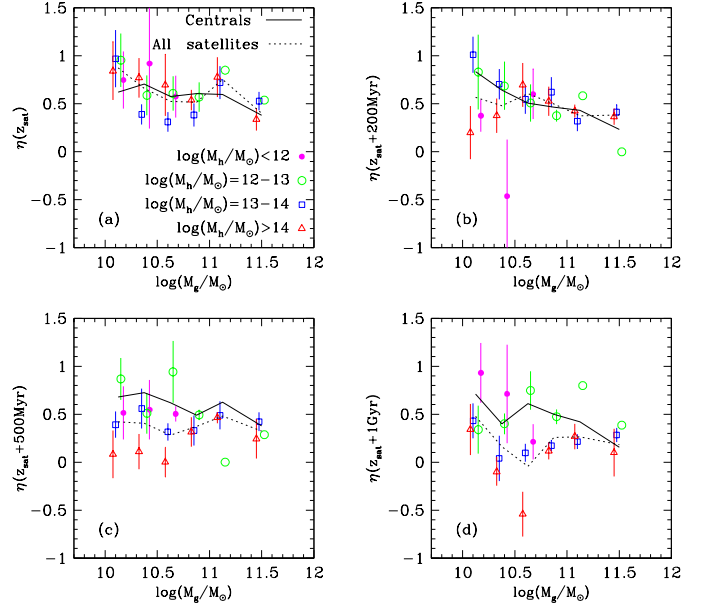}
}
\caption{
Accretion rate against galaxy mass in bins of 0.25 dex. The top left panel
shows the accretion rate at $z_{\rm sat}$, the epoch at which the galaxy becomes a
satellite. The other panels show the accretion rate 200 Myr, 500 Myr and 1
Gyr after $z_{\rm sat}$. Accretion rate is defined as, 
$\eta=(\Delta M_{\rm acc} / M_g)$ $(H^{-1} / t)$
 where $M_{\rm acc}$ is the mass accreted in time $t$, 
$H^{-1}$ is evaluated at the centre of the time interval, and $M_g$ is the
mass of the galaxy at $z=0$. Lines show the average value for all satellites
(dotted) and matched central galaxies (solid) in each galaxy mass bin. Points
with error bars show the average value and uncertainty in the mean in four bins
of halo mass, as indicated in the legend.
}
\label{fig:sub6a}
\end{figure}

\begin{figure}
\centerline{
\epsfxsize=3.5truein
\epsfbox{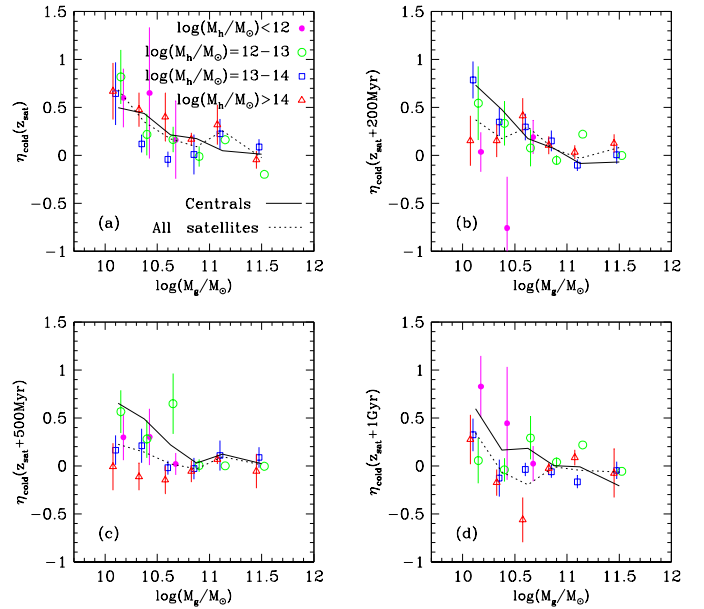}
}
\caption{
Cold accretion rate vs galaxy mass. This figure is analogous to Figure
\ref{fig:sub6a}, but showing only cold accretion.
}
\label{fig:sub6b}
\end{figure}

In our preceding analysis, we examined the mass accreted from
$z_{\rm sat}$ to the present time and found differences in the total
amount of accretion between central and satellite galaxies. To examine
the timescale over which these differences develop, we define a
dimensionless accretion rate as follows:
\begin{equation}
\eta = \frac{\Delta M_{\rm acc}}{M_g}\frac{H^{-1}}{\Delta t},
\end{equation}
where $\Delta M_{\rm acc}$ is the mass accreted in time interval
$\Delta t$, $H$ is the Hubble parameter evaluated at the centre of the
time interval, and $M_g$ is the mass of the galaxy at $z=0$. For each
epoch, we take a small time interval around the central epoch and
calculate the accretion rate using the mass accreted, $\Delta M_{\rm
acc}$ and the time interval $\Delta$t between those epochs. The
precise interval $\Delta$t depends on the simulation outputs available
near $z_{\rm sat}$, but typically $\Delta$t $\sim$ 300 Myr.

Figure \ref{fig:sub6a} compares the accretion rate of satellite
galaxies to the accretion rate of central galaxies in the control
sample at four different epochs. Panel (a) shows the accretion rate at
$z_{sat}$ when the galaxy becomes a satellite, while panels (b), (c)
and (d) show the accretion rates 200 Myr after $z_{\rm sat}$, 500 Myr
after $z_{\rm sat}$ and 1 Gyr after $z_{\rm sat}$, respectively. The
median $z_{\rm sat}$ values in the four halo mass bins are 0.19, 0.55, 
0.59 and 0.99 in increasing order of halo mass. The
dotted and solid curves, averaged over all satellites and matched
centrals in each galaxy mass bin, afford the best statistics for
comparison. At $z_{\rm sat}$ and $z_{\rm sat}$+200 Myr, the
dimensionless accretion rates of satellite and central galaxies are
similar. After 500 Myr, the accretion rates of satellites are
systematically lower, and the difference increases at 1 Gyr. Only the
lowest mass satellites have significantly non-zero accretion rates at
$z_{\rm sat}$ + 1 Gyr. Points with statistical error bars show the
accretion rates in bins of halo mass. The drop in satellite accretion
rates develops most clearly in the highest mass halo, followed by the
next highest mass bin. By $z_{\rm sat}$ + 1 Gyr, some galaxies in the
highest mass halo are losing mass. While this analysis would certainly
benefit from better statistics (i.e. a larger simulation volume), the
trends with time and halo mass are continuous. They indicate that
satellite accretion shuts off gradually, over a timescale of 0.5-1
Gyr, with low mass galaxies in high mass haloes showing the strongest
effect.

Figure \ref{fig:sub6b} repeats this analysis for cold accretion
only. Here the accretion rates are close to zero for all galaxies with
$M_{\rm g}$ above $\sim 10^{10.7} M_{\odot}$, both central and
satellite, at all the epochs shown. At lower $M_{\rm g}$, the
accretion rates are still significantly reduced relative to Figure
\ref{fig:sub6a}, but the differences between central and satellite
galaxies and the dependence on halo mass are similar. It is
particularly notable that low mass satellites in the
$10^{13}$-$10^{14} M_{\odot}$ halo mass bin have positive cold
accretion rates at $z_{\rm sat}$ + 200 Myr and $z_{\rm sat}$ + 500
Myr. Note, however, that the most massive haloes in the bin are $\sim 3
\times 10^{13}$M$_{\odot}$. Even satellites in the $3 \times
10^{14}$M$_{\odot}$ halo show cold accretion at $z_{\rm sat}$ and, to
a lesser extent, at $z_{\rm sat}$ + 200 Myr.

\begin{figure}
\centerline{
\epsfxsize=3.5truein
\epsfbox{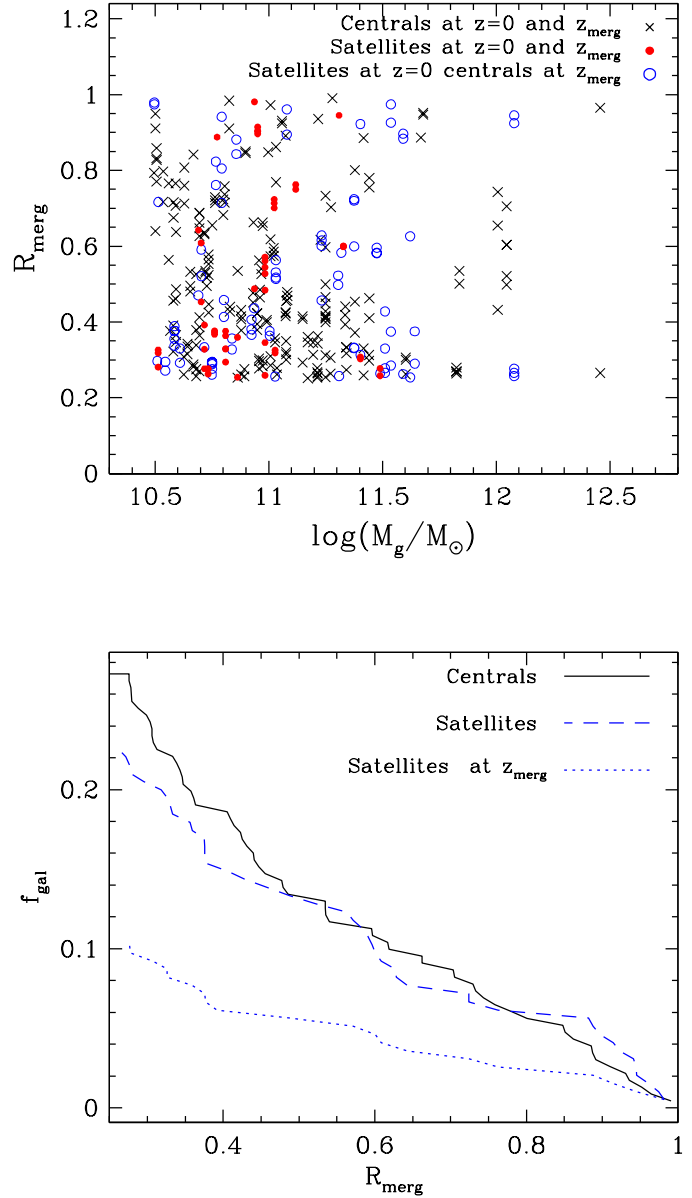}
}
\caption{
\textit{(Top)} Merger events after $z=1$ with mass ratio $R_{\rm merg}$ $\ge$ 0.25, versus
galaxy baryonic mass at $z=0$. Central galaxies, $z=0$ satellites that are centrals at the merger epoch z$_{\rm merg}$ 
and galaxies that are satellites at z$_{\rm merg}$ are shown
separately, as indicated in the legend.
\textit{(Bottom)} Fraction of central galaxies (solid line), $z=0$ satellites
(dashed line) and galaxies that are satellites at $z_{\rm merg}$ (dotted line)
that receive a merger above a certain $R_{\rm merg}$. 
} 
\label{fig:mergers2}
\end{figure}

\begin{figure}
\centerline{
\epsfxsize=3.5truein
\epsfbox{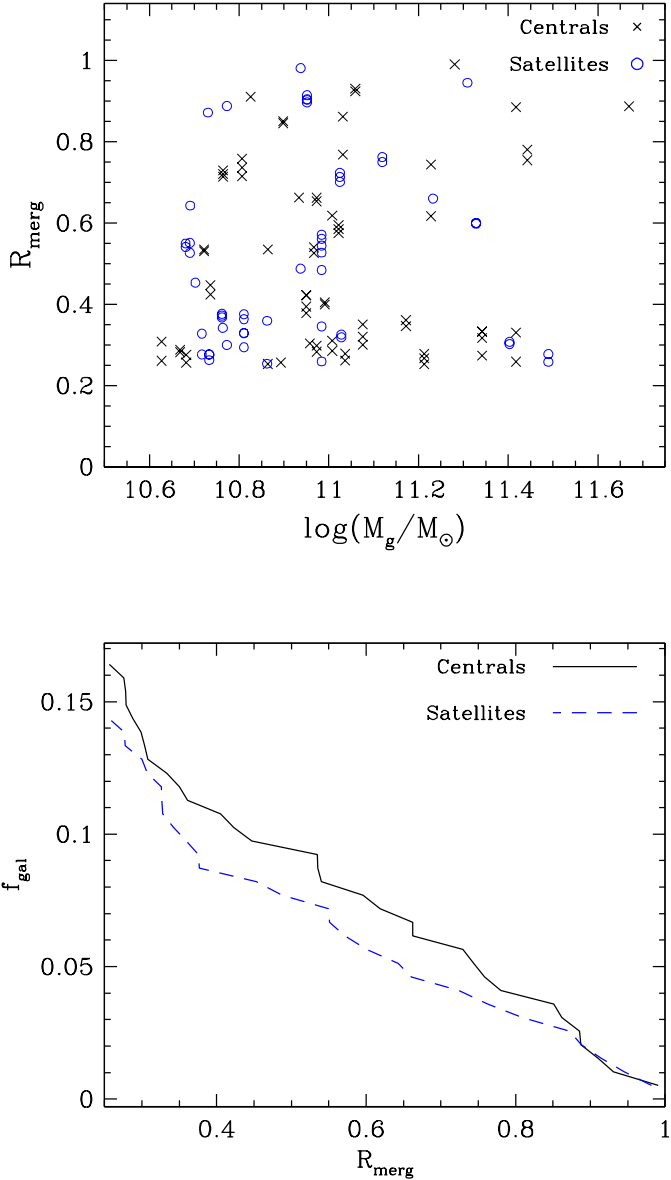}
}
\caption{
\textit{(Top)} Merger mass ratio $R_{\rm merg}$ vs $z=0$ galaxy baryonic mass for
mergers received by satellite galaxies after $z_{\rm sat}$ compared to the
control sample of central galaxies over the same time period. 
\textit{(Bottom)} Fraction of satellite galaxies (dashed line) and central
galaxies in the control sample (solid line) that receive a merger above a
certain $R_{\rm merg}$. 
}
\label{fig:mergers1}
\end{figure}
Besides smooth accretion of gas, galaxies also grow by receiving
mergers from lower mass galaxies. Since we do not resolve galaxies
below $6 \times 10^9$M$_{\odot}$, we restrict our merger analysis to
systems in which the primary galaxy is more massive than $3 \times
10^{10} M_{\odot}$ and the merger mass ratio is $R_{\rm merg} \ge
0.25$. As we go to higher redshifts, the galaxy masses decrease,
causing the number of resolved mergers to decrease. Hence, we restrict
our merger analysis to mergers between $z=1$ and $z=0$. The top panel
of Figure \ref{fig:mergers2} shows $R_{\rm merg}$ for all merger
events between $z=1$ and $z=0$ as a function of the $z=0$ baryonic
mass of the galaxy, differentiating between the central/satellite
status of the primary galaxy at $z_{\rm merg}$ (the epoch of the
merger) and at $z=0$. We see mergers with a wide range of mass ratios
for satellites as well as centrals. A small fraction of mergers
($\sim$10\%) are between objects of similar mass ($R_{\rm merg} \ge
0.9$). More than four-fifths of merger events involve a satellite
galaxy merging with the central galaxy of its parent halo. About
one-third of these galaxies go on to become satellites by $z=0$,
following a merger of their parent halo with a halo hosting a more
massive galaxy. However, it is notable that close to one-fifth of the
merger events in our simulation are mergers between two satellite
galaxies.

The bottom panel of Figure \ref{fig:mergers2} compares the fraction of
central and satellite galaxies that receive mergers between $z=1$ and
$z=0$ above a certain merger ratio, $R_{\rm merg}$. One-fourth of the
galaxies in the simulation (above $3 \times 10^{10}M_{\odot}$)
experience at least one merger with R$_{\rm merg} \ge 0.25$ between
$z=1$ and $z=0$. This fraction is similar for central galaxies and for
$z=0$ satellites and the $R_{\rm merg}$ distribution is also similar,
as one can see by comparing the solid and dashed lines. However, the
fraction of satellites that receive mergers after $z_{\rm sat}$ is a
factor of 2-3 lower (dotted line). Galaxies that receive mergers often
receive more than one merger, so the accounting in the top and bottom
panels is somewhat different.

Figure \ref{fig:mergers2} adopts the somewhat arbitrary redshift
interval $z=1$ to $z=0$, and the mass distributions of the central and
satellite galaxies are systematically different. To fairly compare the
merger activity of central galaxies and post-$z_{\rm sat}$ satellites,
we return to the control sample used for our accretion
comparisons. The top panel of Figure \ref{fig:mergers1} shows the
merger ratio for mergers (no longer restricted to $z\le1$) 
with $R_{\rm merg} \ge 0.25$ received by
satellite galaxies and by the control sample of central galaxies after
$z_{\rm sat}$ of the matched satellite. Only 14\% of satellite
galaxies and 16\% of central galaxies in the control sample receive
mergers after $z_{\rm sat}$. The distribution of mass ratios is broad
for both populations. Our sample is not large enough to study the
merger properties of satellite galaxies as a function of halo
mass. The bottom panel of Figure \ref{fig:mergers1} shows the fraction
of central and satellite galaxies that receive a merger above a given
$R_{\rm merg}$ after $z_{\rm sat}$ (in contrast to Figure
\ref{fig:mergers2}, which shows all mergers since $z=1$ and includes
all central galaxies). The probability of a satellite galaxy receiving a merger above a
certain mass ratio is only $\sim$2\% lower than that of a central galaxy 
of similar mass, across a wide range of
merger mass ratios.

\section{DISCUSSION}

The central galaxies of haloes form a distinct population in this
simulation, as anticipated by semi-analytic models
\citep{white91,kauffmann93,cole94} and as found in previous numerical
studies \citep{berlind03,zheng05}. In the great majority of FOF haloes,
the most massive galaxy lies at the bottom of the dark matter
potential well and close to the centre of mass. In more than
two-thirds of the FOF haloes, the baryonic mass (stars + cold gas) of
the central galaxy is 2-5 times higher than that of the most massive
satellite. Only a small fraction of FOF haloes host two galaxies of
similar mass. Because of its limited volume, our simulation has only
one cluster mass halo ($M$ $\sim$ $3 \times 10^{14} M_{\odot}$). The
other haloes massive enough to host resolved satellites range from
$\sim 3 \times 10^{11} M_{\odot}$ to $\sim 3 \times 10^{13}
M_{\odot}$, while haloes down to $\sim 10^{11}M_{\odot}$ host resolved
central galaxies.

The distinction between central and satellite galaxies in our
simulation is weaker than expected in a simple picture where only
central galaxies accrete mass and ``receive'' mergers of less massive
systems. Galaxies that are satellites at $z=0$ assembled their mass
earlier, on average, than present day central galaxies of the same
baryonic mass, and satellites of more massive haloes have
systematically higher assembly redshifts. However, to a large extent
the baryonic mass of a galaxy is a better predictor of its accretion
and merger history than its central or satellite status, with the
exception of galaxies in the cluster mass halo. In part, the dominance
of baryonic mass over satellite status just reflects the fact that
many satellites were central galaxies for most of their history,
merging into larger haloes only at low redshift. However, we find that
simulated galaxies continue to accrete gas and, in some cases, merge
with lower mass objects even after they become satellites. In essence,
galaxies that enter the virial radius of a larger halo may remain the
``central'' galaxies of their own dark matter substructure, and they
lose their central status only gradually. Alternative halo definitions
(e.g., higher threshold density or spherical overdensity identification)
could yield somewhat sharper distinctions between central and satellite
galaxies, but we do not expect them to change this basic result.

While galaxies continue to accrete gas after becoming satellites, they
accrete less than a ``control'' sample of central galaxies with
similar baryonic masses measured over the same time period. At the
time they enter a larger halo, satellites have similar accretion rates
to central galaxies of the same mass, and the accretion rate declines
towards zero over the next 0.5-1 Gyr. Lower mass satellites in the
cluster mass halo begin to lose mass, and by $z=0$ they are typically
10-20\% less massive than they were when they became satellites. When
we compare satellite galaxies to central galaxies of the same baryonic
mass, we find large systematic differences if, and only if, the mass of
the satellite's host halo is much larger than the typical halo mass for the 
corresponding central galaxies.

As in \cite{keres05}, we find that galaxies below $M_g \sim 3 \times
10^{10} M_{\odot}$ accrete predominantly through ``cold mode'', while
higher mass galaxies grow mainly through ``hot mode'' accretion of
virial temperature gas. This transition is similar for central and
satellite galaxies. Consequently, our general conclusions would still
hold if we assumed that a galaxy's true accretion rate was close to
its cold (rather than total) accretion rate, for either the physical
or numerical reasons discussed in \S3, except that the accretion rates
of massive galaxies would become very low. Regardless of whether we 
consider total accretion or cold accretion only, these simulations
exhibit natural ``downsizing'' in which accretion and star formation
shift towards lower mass galaxies at later times \citep{keres05,keresp}.

Comparing present day central galaxies and present day satellites,
above a mass threshold of $M_{\rm gal} = 3 \times 10^{10} M_{\odot}$,
we find that only a slightly larger fraction of centrals (27\%
vs. 22\%) have received mergers with mass ratio $R_{\rm merg} \ge
0.25$ since $z=1$. Roughly half of the satellite mergers occurred
after the larger parent had become a satellite, and half occurred
while it was still a central object. The mass ratio distributions are
similar in all cases. There are more central galaxies than satellites
and only about one in five resolved merger events involves two
satellites. However, if we compare satellites to a control sample of
centrals as we did for accretion, we find similar fractions
experiencing resolved, $R_{\rm merg} \ge 0.25$ mergers during the
course of the simulation, with similar $R_{\rm merg}$
distributions. In summary, most galaxies have quiescent assembly
histories at $z<1$, and the merger rates of central and satellite
galaxies are not radically different. The most massive galaxies do
have more complex merger histories \citep{maller06}, and these are
usually central galaxies of high mass haloes.

While we have focused on mass growth rather than star formation, the
two are closely linked in our simulation, with star formation
generally following gas accretion after a short delay. The fact that
galaxies experience continuing gas accretion for 0.5-1 Gyr after
becoming satellites means that satellites will be systematically bluer
than predicted in semi-analytic models that assume no accretion onto
satellites (``strangulation''). There are numerous indications that
including the continuing satellite accretion predicted here would
improve the match between semi-analytic models and
observations. \cite{weinemann06} find that the \cite{croton06}
semi-analytic model predicts an excessive fraction of red satellites
relative to the satellite population in groups identified from the
Sloan Digital Sky Survey (SDSS). The discrepancy is largest for lower
mass groups, which have much higher blue fractions than predicted,
consistent with our finding that the central-satellite distinction
becomes sharper for satellites in high mass haloes. \cite{cli07} show
that the pairwise velocity distributions predicted by the
\cite{kang05} and \cite{croton06} models can be brought into agreement
with SDSS measurements by reducing the fraction of faint red
galaxies. \cite{coil08} find similar trends at $z \sim 1$, showing
that the \cite{croton06} model predicts excessive clustering of red
galaxies and insufficient clustering of blue galaxies relative to the
DEEP2 Galaxy Redshift Survey, especially on small scales. Most recently, 
\cite{weinemann08} show that the colours of SDSS satellites can be reproduced
if their star formation slows over a timescale of 2-3 Gyr after $z_{\rm sat}$, 
somewhat larger than the timescale for accretion shut off found here.

Other observations indicate that the colours of satellite galaxies
depend systematically on halo mass, as we would expect based on the
assembly redshift trends in Figures 6 and 8. In a sample of clusters
taken from the Red-Sequence Cluster Survey, \cite{gilbank07} find that
the red fraction in clusters increases with richness, indicating an
earlier truncation of star formation in higher mass
haloes. \cite{hansen07} find a similar result using data from the
SDSS. Finally, X-ray studies of clusters indicate that massive
satellites retain local concentrations of hot gas
\citep{sun07,jeltema08}, providing further evidence that the physical
transition from ``central galaxy'' to ``satellite'' is a gradual one.
Further discussions of the impact of the halo environment on satellite
galaxy properties, focused on the effects of ram pressure stripping, are
given by \cite{mccarthy08} and \cite{font08}.

Our small simulation volume limits our statistical power, especially
for massive haloes. In addition, there are indications that the TreeSPH
code used here predicts too much hot gas accretion, and that the
entropy conserving SPH formulation of \cite{springel02} yields more
accurate results \citep{springel02,keresp,keres07}. \cite{keresp} have
recently analysed a simulation of a 50$h^{-1}$Mpc cube evolved with
the entropy conserving SPH code GADGET-2 \citep{springel01}, with a
similar mass resolution to the simulation analysed here. While they
focus mainly on the relative importance of cold and hot accretion in
galaxy growth, they also investigate central and satellite accretion,
and confirm our main result: satellites experience significant amounts
of continuing accretion, and the distinction between central and
satellite growth rates is smaller at high galaxy masses and (a point
not investigated here) at high redshift. Both our simulation and the
\cite{keresp} simulation overpredict the observed galaxy baryonic mass
function, probably because they do not include galactic winds that
eject accreted material from low and intermediate mass galaxies. Winds
will change the relation between accretion rates and star formation
rates, and they will affect the accretion rates themselves by altering
the local gas supply. However, we do not expect them to alter our
qualitative conclusions: the distinction between central and satellite galaxies is
fundamental to understanding the galaxy population and its evolution,
but differences in their growth rates and hence their physical
properties are smooth, not sharp.

\section*{ACKNOWLEDGMENTS}
We thank Mark Fardal and Ari Maller for useful discussions and technical
assistance. This work was supported by NASA Grants NAG5-13308
and NNG04GK68G and by an Ohio State University Graduate Fellowship.

%\singlespace
\bibliographystyle{mn2e}

\end{document}